\RequirePackage{fix-cm}
\documentclass[smallextended]{svjour3}       
\smartqed  
\usepackage{graphicx}
\usepackage{booktabs}       
\usepackage{hyperref}
\usepackage{url}
\usepackage{amsmath}
\usepackage{color}
\usepackage{tikz}

\def\hyphenateAndTtWholeString #1{\xHyphenate#1$\wholeString\unskip}

\def\xHyphenate#1#2\wholeString {\if#1$%
    \else\transform{#1}%
    \takeTheRest#2\ofTheString\fi}

\def\takeTheRest#1\ofTheString\fi
{\fi \xHyphenate#1\wholeString}

\usepackage{pifont}
\newcommand{\gcmark}{\textcolor{green}{\ding{51}}}%
\newcommand{\rxmark}{\textcolor{red}{\ding{55}}}%
\newcommand{\varga}{\cite{varga2019no}}

\def\@fnsymbol#1{\ensuremath{\ifcase#1\or *\or \dagger\or \ddagger\or
   \mathsection\or \mathparagraph\or \|\or **\or \dagger\dagger
   \or \ddagger\ddagger \else\@ctrerr\fi}}
\newcommand{\ssymbol}[1]{$^{\@fnsymbol{#1}}$}

\def\transform#1{\url{#1}\hskip 0pt plus 1pt}

\def\urlx #1{\href{#1}{\hyphenateAndTtWholeString{#1}}}

\begin{document}

\title{Comment on ``No-Reference Video Quality Assessment Based on the Temporal Pooling of Deep Features"}

\titlerunning{Comment on ``NR-VQA based on Temporal Pooling of Deep Features"}        

\author{Franz G\"otz-Hahn, Vlad Hosu, Dietmar Saupe}


\institute{F. G\"otz-Hahn \at
              \email{franz.hahn@uni.kn}           
           \and
           V. Hosu \at
              \email{vlad.hosu@uni.kn}           
           \and
           D. Saupe \at
              \email{dietmar.saupe@uni.kn}           
              }

\date{Received: date / Accepted: date}

\maketitle

\begin{abstract}
In Neural Processing Letters 50,3 (2019) a machine learning approach to blind video quality assessment was proposed \cite{varga2019no}. It is based on temporal pooling of features of video frames, taken from the last pooling layer of deep convolutional neural networks. The method was validated on two established benchmark datasets and gave results far better than the previous state-of-the-art. In this letter we report the results from our careful reimplementations. The performance results, claimed in the paper, cannot be reached, and are even below the state-of-the-art by a large margin. We show that the originally reported wrong performance results are a consequence of two cases of data leakage. Information from outside the training dataset was used in the fine-tuning stage and in the model evaluation. 
\end{abstract}

\section{Introduction}
\label{sec:intro}
For the design of video processing methods and their practical use, objective video quality assessment (VQA) is required. This refers to the algorithmic estimation of subjective video quality, as experienced by human observers. In order to develop such algorithms, benchmark datasets have been created that contain video sequences together with quality attributes. These quality labels usually are mean opinion scores (MOS) from lab-based or crowdsourced user studies. They serve as ground truth data for model evaluation, as well as for training/validation of machine learning approaches.

VQA comes in different flavors, most notably as the so-called full-reference (FR) VQA and as no-reference (NR, or blind) VQA. For FR-VQA an original pristine video is compared to a distorted version of the same video, and the quality difference between the two is evaluated. In this note, we discuss the work of Varga \cite{varga2019no}, which proposes a machine learning approach for blind VQA, i.e., where the only input to the VQA algorithm is the distorted test video, the visual quality of which is to be assessed. For an introduction to video quality assessment refer to \cite{chen2014qos}.

Deep convolutional neural networks (DCNN) have seen increased use as tools for feature extraction for a variety of perceptual tasks in recent years~\cite{hii2017multigap,bianco2018use,gao2018blind,zhang2018unreasonable,hosu2019effective}. In \cite{varga2019no}, the author proposed an approach to frame-level feature extraction for VQA. In a nutshell, it works as follows. DCNN architectures, pre-trained in an object classification task, such as Inception-ResNet-v2 or Inception-v3, are fine-tuned in a five-class classification task. The inputs in this fine-tuning process are individual video frames, and the target classes represent intervals of the source video's quality MOS. After fine-tuning, video frames are passed through the network sequentially, and the activations of the last pooling layer are extracted and saved as their feature representations. In order to obtain video-level features, frame-level features are aggregated by performing average, median, minimum, or maximum pooling. Finally, these aggregated video feature vectors serve as input to a support vector regressor (SVR). The author compared a variety of SVR kernel functions for the different aggregation methods. Note that the approach is very similar to related works in the image and aesthetics quality assessment domains, e.g., with \cite{gao2018blind} and \cite{hii2017multigap}. In these works, pre-trained networks were used with or without fine-tuning to extract features and predict perceptual attributes.


For training, validation, and testing of the deep network and the SVR, the well-established video dataset KoNViD-1k was used \cite{hosu2017konstanz}. The best performance was achieved using an Inception-V3 network architecture as a feature extractor, average pooling the individual frame-level features, and with the SVR being trained using a radial basis function (RBF) kernel. A common performance metric reported for VQA algorithms is the correlation between the model predictions and the ground truth MOS. In \cite{varga2019no}, the peak average performance on test sets from KoNViD-1k was given by a  Pearson linear correlation coefficient (PLCC) of 0.853 and a  Spearman rank-order correlation coefficient (SROCC) of 0.849. In the paper, another dataset (LIVE-VQA) was also used, however, for brevity and simplicity, we focus on the former one in our discussion here.

The previous best-reported performance on KoNViD-1k was achieved by TLVQM~\cite{korhonen2019two}, with a 0.77 PLCC and 0.78 SROCC, respectively. The improvement in performance of 0.08 PLCC and 0.07 SROCC is substantial, considering the usually incremental improvements in the field.

The author of \cite{varga2019no} provided code for his method on his personal GitHub repository\footnote{\urlx{https://github.com/Skythianos/No-Reference-Video-Quality-Assessment-Based-on-the-Temporal-Pooling-of-Deep-Features}, available since July 29, 2019, and revised on August 9, 2019.}. Based on this code, we reimplemented and tested the method, as described in the paper. It did not produce the results as claimed. However, we also succeeded in reverse-engineering an implementation that did give the numbers as in the paper. The key observation for this task was some data-leaking in the fine-tuning stage that became already apparent from the first version of the author's code on GitHub. The author was made aware of this data leak after the publication and subsequently corrected the mistake\footnote{\urlx{https://github.com/Skythianos/No-Reference-Video-Quality-Assessment-Based-on-the-Temporal-Pooling-of-Deep-Features/issues/2}}. Extrapolating this mistake to the SVR in the second stage of the method finally produced the numbers as published in the paper.


In this communication we will share and discuss the correct results for the approach, describe the reverse-engineering process, and show the mistakes that have likely resulted in the incorrect published performance numbers. The complete code necessary to reproduce the results in this report is available online.\footnote{
See \urlx{https://github.com/FranzHahn/NPL-50-3-2595-2608-Correction}. There we also included links to MATLAB workspace binaries containing trained networks, extracted features, as well as non-aggregated results.}
From our analysis we conclude the following:
\begin{enumerate}
    \item The method as described in \cite{varga2019no} does not yield the performance as claimed. On KoNViD-1k, the SROCC on test sets is only $0.69 \pm 0.04$ instead of 0.85.
    \item We show that the discrepancy between these results can be attributed to a twofold data leakage. First, the validation set, and then the test set was not properly separated from the training set.
    \item Na\"ive fine-tuning of Inception-style networks using either the classification method described in \cite{varga2019no} or regressing mean opinion scores in an end-to-end fashion is not a promising solution for VQA.
\end{enumerate}

\section{Fine-Tuning}
\label{sec:finetuning}

In \cite{varga2019no}, the author describes the fine-tuning process as follows. A pre-trained Inception-style network is modified, such that the final fully-connected (FC) softmax layer is replaced with a 5-way FC softmax layer, using the Xavier weights initialization. The output of the neurons in this layer correspond to the five intervals that contain the video's mean opinion score (MOS). Note that the network inputs are video frames. Concretely, the class $C(v[i])$ for the $i$th frame of video $v$ as an input to the network is assigned as:
\begin{equation}
C(v[i]) = 
\begin{cases}
\mathrm{VeryGood} & \text{if } 4.2 \leq \mathrm{MOS}({v}) \leq 5.0,\\
\mathrm{Good} & \text{if } 3.4 < \mathrm{MOS}({v}) \le 4.2,\\
\mathrm{Mediocre} & \text{if } 2.6 < \mathrm{MOS}({v}) \le 3.4,\\
\mathrm{Poor} & \text{if } 1.8 < \mathrm{MOS}({v}) \le 2.6,\\
\mathrm{VeryPoor} & \text{if } 1.0 < \mathrm{MOS}({v}) \le 1.8.
\end{cases}
\end{equation}

Fine-tuning was performed using stochastic gradient descent (SGD) with momentum $\beta = 0.9$ and an initial learning rate $\alpha = 10^{-4}$. The author states that the rate was divided by 10 when the validation loss stopped decreasing during training, although the code available online does not do this.

\begin{figure*}[t!]
    \centering
    \includegraphics[width=0.95\columnwidth]{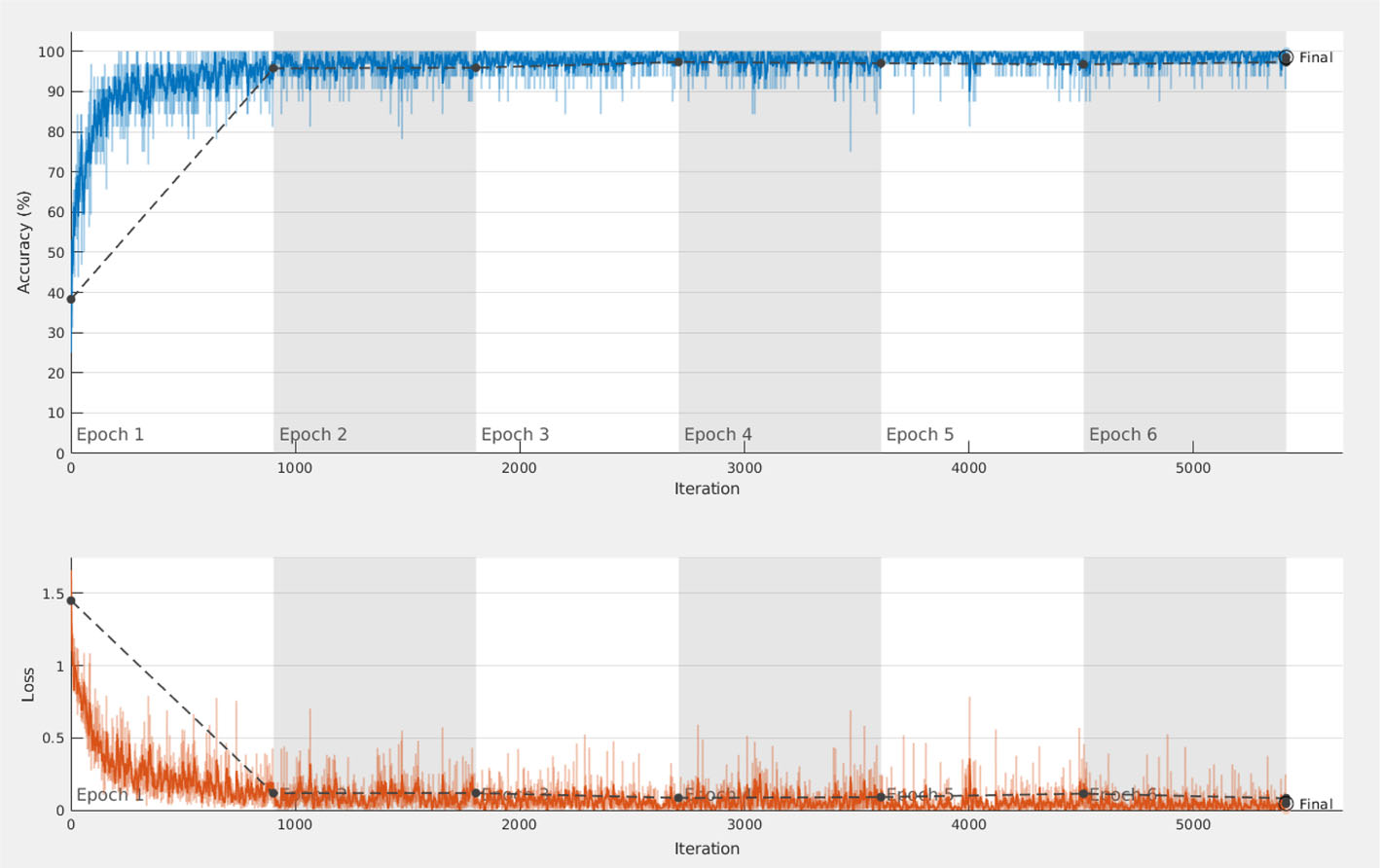}
    \caption{The training progress during fine-tuning as reported in \cite{varga2019no}. The blue lines show smoothed and per iteration training accuracies in dark and light color variants, respectively. Similarly, the orange lines depict smoothed and per iteration training losses in dark and light color variants, respectively. The dashed dark gray lines linearly connect the validation accuracies and losses indicated by the dark gray circle markers.}
    \label{fig:fine-tune-repro}
\end{figure*}


The KoNViD-1k dataset was used for evaluation of this approach. It consists of 1,200 video sequences with accompanying MOS values. 960 videos were randomly chosen for training, splitting the dataset 4:1. More precisely, 20\% of the frames of the 960 videos were randomly selected constituting the combined training and validation set for the fine-tuning and feature learning. The remaining 240 videos were reserved as a test set and not used during the fine-tuning step.

The set of extracted frames was further divided into a training and validation set. Although the paper does not specify what ratio was used (training to validation set size), it can be assumed that the ratio was 3:1, as an overall 3:1:1 ratio between training, validation, and test sets is common in deep learning.\footnote{Different online versions of the author's code use different secondary splits. Both 2:1 and 3:1 have been employed in different versions.} 

\begin{figure*}[t!]
    \centering
    \includegraphics[width=0.95\columnwidth]{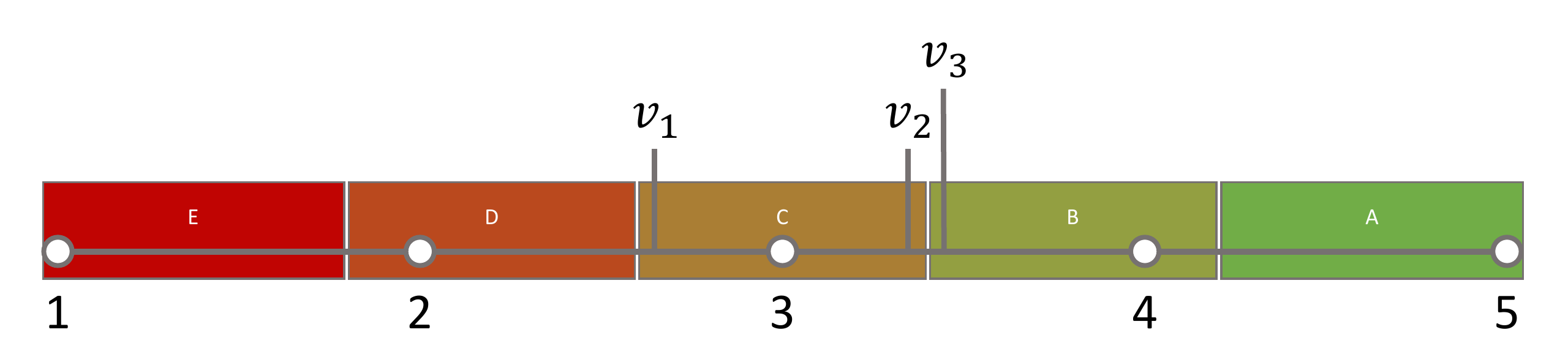}
    \caption{A diagram of the MOS scale with fine-tuning class labels derived from a video's MOS value. The classes are given as colored rectangles. Given three videos $v_1$, $v_2$, and $v_3$ at adjacent class boundaries, the difficulty of the classification task becomes apparent. The perceptual attributes of $v_2$ and $v_3$ are likely to be very similar, but they are grouped into different classes. Conversely, $v_1$ and $v_2$ will share less perceptual similarity than the previous pair, but they are grouped into the same class. Furthermore, the difference in perceptual similarity between the pairs ${v_1, v_2}$ and ${v_1, v_3}$ is likely small, but the model has to distinguish it somehow.}
    \label{fig:class_complexity}
\end{figure*}

As a result of the training for the classification task, the author reported in \cite{varga2019no} a classification accuracy on the validation set after fine-tuning that is higher than 95\% . Unfortunately, this high validation accuracy is not achievable when implementing the approach as described. In fact, to an observer familiar with machine learning, the author's fine-tuning training progress plot (reproduced in Figure~\ref{fig:fine-tune-repro}) raises two questions:
\begin{enumerate}
    \item  The quick increase of both the training and validation accuracies may be possible for such a training process, however, considering the broadness and complexity of the classes, it seems unlikely. At class boundaries, the classification task is hard as illustrated in Figure~\ref{fig:class_complexity}. Based on perceptual information alone, a human would be hard-pressed to perform the classification correctly. It is very unlikely that a classification accuracy of above 95\% for the validation set is achievable in such a difficult scenario.
    \item Complex DCNNs, trained on small datasets, like the one used in this work, eventually overfit if training keeps going on long enough. The validation set is meant as a tool to detect overfitting and, therefore, as a criterion to stop training. Overfitting can be detected by comparing the change in validation set performance. Conventionally, when overfitting occurs, validation set performance starts dropping as the training set performance is steadily rising. However, in this plot there is no such \textit{noticeable} drop in the validation set accuracy. 
    Consequently, more training should have been performed to make full use of an independent validation set.
\end{enumerate}

\begin{figure*}[t!]
    \centering
    \includegraphics[width=0.95\columnwidth]{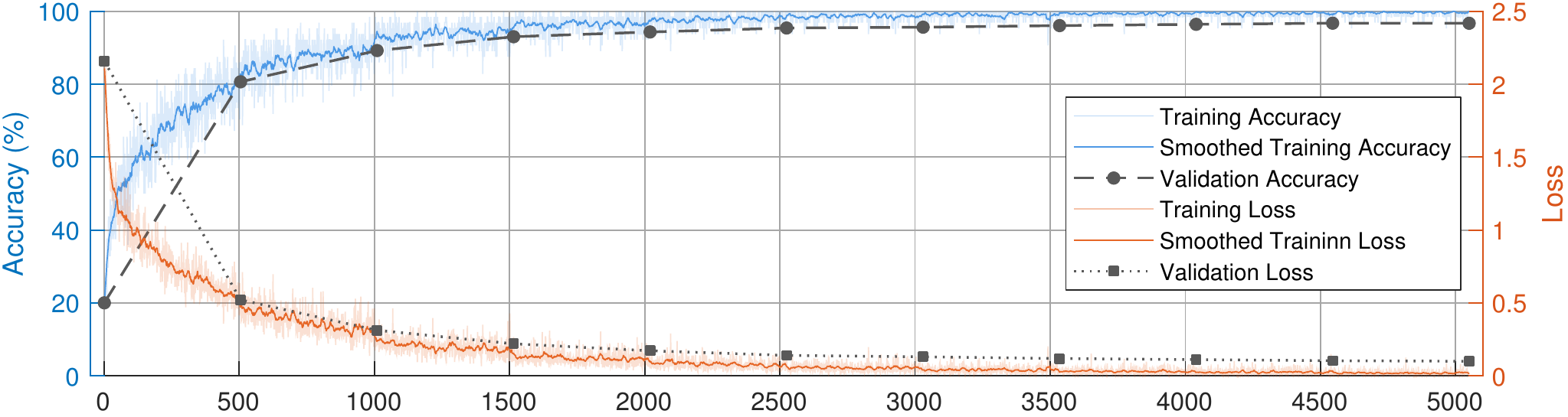} \\
    \includegraphics[width=0.95\columnwidth]{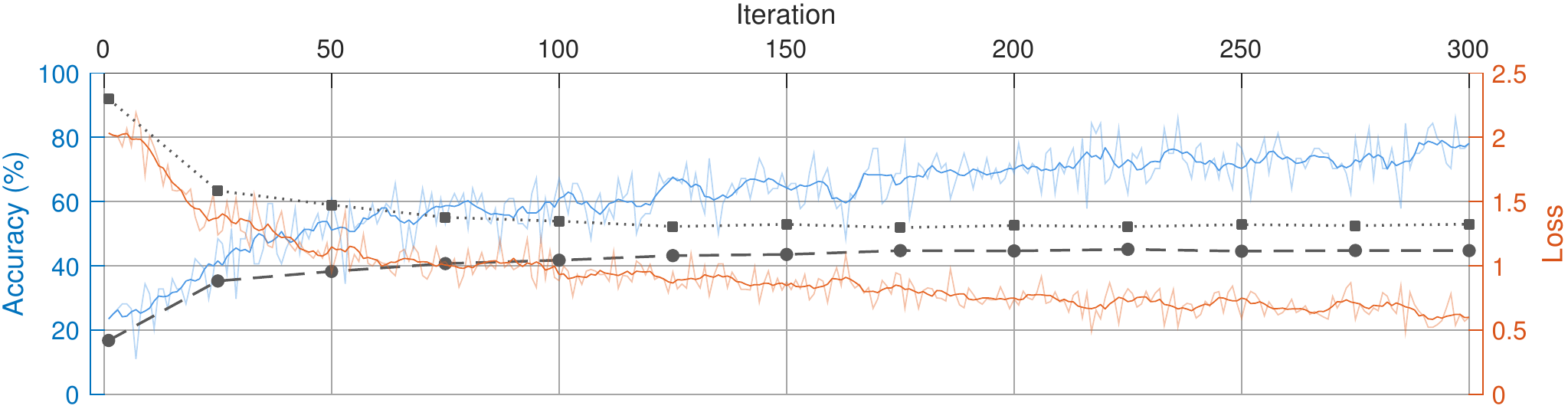}
    \caption{Comparison of reimplementations of the fine-tuning procedure. The top figure depicts the training progress of a fine-tuning procedure with data leakage, while the bottom figure shows the training progress of a fine-tuning procedure without data leakage.}
    \label{fig:fine-tune}
\end{figure*}

Figure~\ref{fig:fine-tune} depicts the training progress of the fine-tuning step of our reimplementation of the author's approach in the upper part and its corrected version below. In order to obtain the plot in the upper part, we had to introduce what is called data leakage~\cite{nisbet2017handbook}. Data leakage can be understood in different ways, but it always describes situations where information sources that are meant to be independent are influencing each other and are, therefore, not independent. This particular form of data leakage can actually be found in an earlier version of the author's public code\footnote{\urlx{https://github.com/Skythianos/No-Reference-Video-Quality-Assessment-Based-on-the-Temporal-Pooling-of-Deep-Features/tree/621f689eae8319be79af80497db55d97637ea213}}, and is therefore likely to have been the cause of this implausible fine-tuning performance. The author was notified of this error in August 2019, as can be seen in the discussion of this problem on the author's code repository issues page\footnote{\urlx{https://github.com/Skythianos/No-Reference-Video-Quality-Assessment-Based-on-the-Temporal-Pooling-of-Deep-Features/issues/2\#issue-475618103}}.

In the author's original implementation, the first selection of 80\% of the videos was for the purpose of fine-tuning the feature extraction network. Then 20\% of the frames from these videos were randomly selected and pooled in a data structure. From this data structure, the program made a random selection for training and validation. Obviously, this causes frames from the same video to end up in both of the sets. This defeats the purpose of the validation set. By sampling the validation set independently from the training set it should give an indication about the generalization power of the model on an independent test set. Since this sampling does not result in an independent set, the validation performance does not indicate what the performance on a test set could be. In fact, it only gives the same information as the training set performance, as the two sets are near identical in content.

Consequently, the model overfits on the training set and this cannot be detected by the validation set that was chosen in \cite{varga2019no}. The fine-tuned model should have poorer performance on an arbitrary set of videos that is independent of the training set as is the case for the test set. From the earlier versions of the author's code as well as from Figure~\ref{fig:fine-tune-repro}, it must be concluded that this data leakage was present in the particular implementation that was used in \cite{varga2019no} to produce the reported results.

Furthermore, some parameters for the training process were poorly chosen. Evaluation of the validation set is conventionally performed once per epoch, where an epoch describes the entire training data being passed through the network once. If the inputs are independent images, e.g., in an object classification problem, this is a reasonable approach. However, in this case, the training set consists of 20\% of all frames from each video selected for training. In the case of a video with 240 frames this results in 48 frames from the same video being passed through the network, before the validation set is being evaluated. Compared to the object classification task on images from above this is comparable to 48 epochs. As mentioned above, the evaluation of the validation set is used to select the best generalizing model. Infrequent validation can lead to poor model selection. Therefore, we evaluated the validation set more frequently in our reimplementation, in order to select the best performing feature extraction model as a basis for further steps. Validation occurred once every 1600 frames in our training procedure, as compared to once every 32,000--33,000 frames in the original implementation. Comparing the two plots in Figure~\ref{fig:fine-tune} we can see that the training procedure shown in the bottom stops at iteration 300. Here, validation loss is stagnating, while training loss keeps decreasing, which is an indicator for overfitting on the training data. However, in the top plot the first validation set evaluation only occurs after 500 iterations. If we were to employ the same validation frequency, we would likely not be able to select a well performing model.

\begin{figure*}[t!]
    \centering
    \includegraphics[width=0.95\columnwidth]{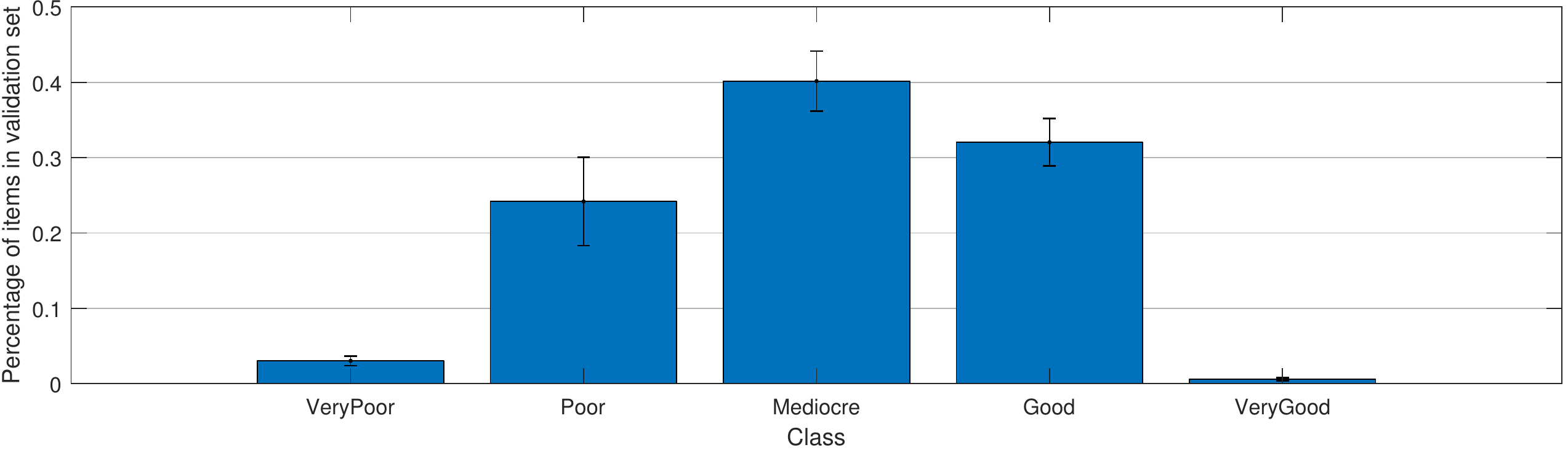}
    \caption{Average distribution of class predictions in percent across the five splits used for the fine-tuning of the feature extraction model. The error bars denote the standard deviation.}
    \label{fig:class_dist}
\end{figure*}

Secondly, the fine-tuning process in itself does not seem to have a big impact. 
Figure~\ref{fig:class_dist} shows the distribution of predicted video classes in the test set averaged over five random splits with the error bars representing the standard deviation. The average peak test accuracy for the classification task across five correctly fine-tuned models is $46.52\%$. The average test set accuracy when predicting the dominant class is $41.08\%$. This $5.44\%$ increase in overall prediction accuracy over predicting the dominant class does not show a large improvement and indicates, that the classification task may not be appropriate. This could be due to the problems with lumping MOS scores into coarse classes as described earlier or a more general problem of Inception-V3 features not being informative about video quality. We investigate the latter in Section~\ref{sec:discussion}. 

\section{Quality Prediction}
\label{sec:qualityprediction}
After the network was fine-tuned for the classification of MOS intervals, the model was used as a feature extractor in \varga. By passing a video frame to the network and extracting the activations of the final pooling layer, a high-level feature representation of the input was obtained. The feature vectors resulting from the frames of a given video sequence were aggregated by either computing the mean, median, minimum, or maximum values, yielding a video-level feature representation. In order to predict the video's quality,  an SVR was trained with different kernel functions. Final results were presented as Pearson linear correlation coefficients (PLCC) and Spearman rank-order correlation coefficients (SROCC) between the predictions and the ground-truth MOS values. In addition, the paper \varga\ provides the performance for the SVR trained on the features extracted from the off-the-shelf Inception-style network without the proposed fine-tuning. 

The best and main result of the paper is the peak performance that was obtained with a fine-tuned Inception-V3 network, feature aggregation using average pooling, and an SVR trained with a radial basis function kernel. With this setup, a PLCC of 0.853 and an SROCC of 0.849 were reported for \mbox{KoNViD-1k}.

The state-of-the-art performance on KoNViD-1k at the time of the publication \varga\ was 0.77 PLCC and 0.78 SROCC \cite{korhonen2019two}. The reported improvement is substantial and surprising. However, the claimed performance is not reproducible. In the following we describe our reimplementation and the true performance achievable with the method of \varga. Furthermore, taking a closer look at the way the SVR was applied, we can explain the faults of the method that caused the dubious performance results in \varga. It is another case of data leakage, this time from the feature learning network into the SVR test sets. This is verified by a reconstruction of this data leakage, which reproduces the results in \varga.


\begin{figure*}[t!]
    \centering
    \includegraphics[width=\columnwidth]{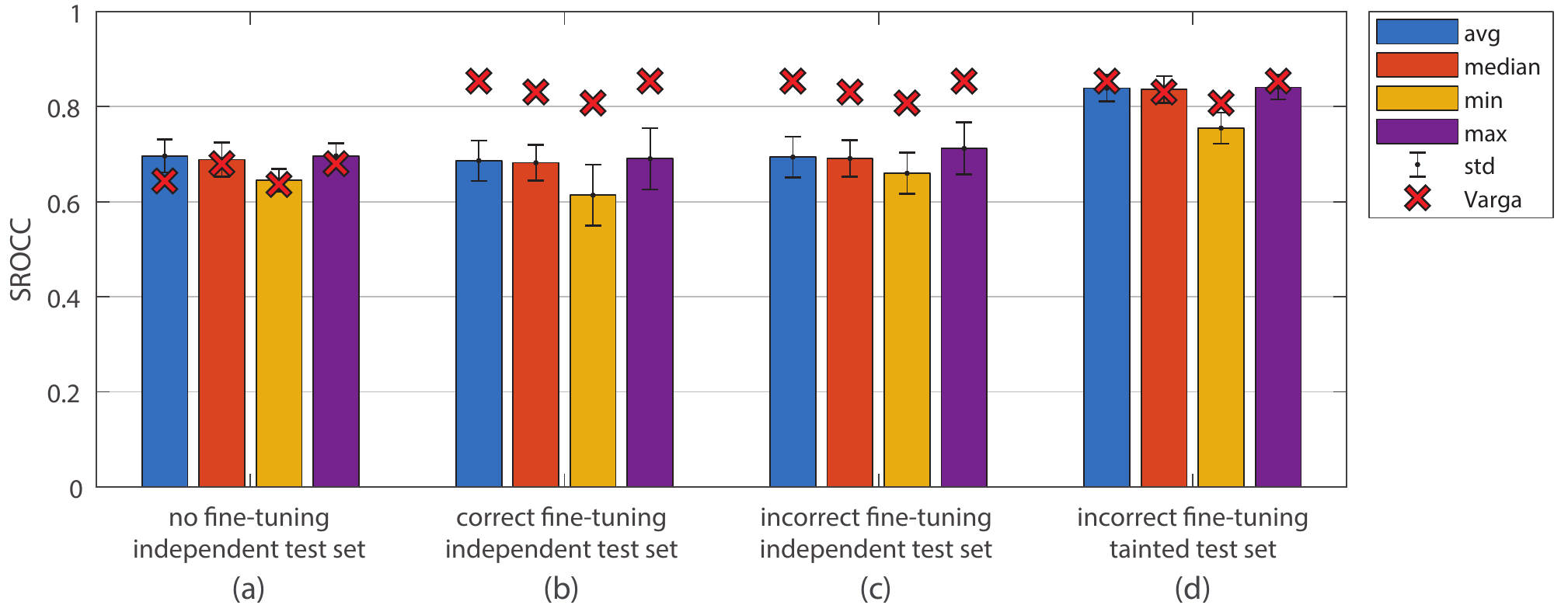}
    \caption{Performance comparison of SVRs trained using different kernel functions from our reimplementation. Chart (a) shows the results when no fine-tuning is used for the feature extraction network. The performance with correctly applied fine-tuning is shown in chart (b), which is also the \textit{true} performance of the approach. Charts (c) and (d) depict the performance when fine-tuning is performed with data leakage. The bars represent average performance of five random training, validation, and test splits. Independent test sets are chosen prior to fine-tuning, and for (d) also tainted test sets are chosen at random before SVR training. The red cross markers represent the corresponding numbers reported by Varga in \varga, as measured from the figures in the paper.}
    \label{fig:results}
\end{figure*}

Figure~\ref{fig:results} (a) shows the average performance of five SVRs trained with a gaussian kernel function without fine-tuning of the feature extraction network. The approximate results reported in \varga\ as measured from the figures in the original paper are shown by the red cross markers, and they match those of our reimplementation. In this case, the fine-tuning data leakage described in the previous section has no effect, as no fine-tuning is employed.

Chart (b) in Figure~\ref{fig:results}, on the other hand, shows the performance of SVRs trained on the same splits but with correctly implemented fine-tuning in the first step. More importantly, the SVRs were trained using only the training and validation set videos that were already used in the fine-tuning process. The test set was not made available at the fine-tuning stage, nor in the training of the SVR model.

We see a vast difference in performance between our reimplementation and the performance numbers reported by the author as denoted by the red crosses. How could that have happened? The differences cannot solely be attributed to the incorrect fine-tuning. Figure~\ref{fig:results} (c) depicts the average performance values of the five SVRs with incorrect fine-tuning evaluated on the independent test sets with little improvements over chart (b). This begs the question of what might have happened in the performance evaluation process in \varga.



The standard practice when training a machine learning regressor is to utilize k-fold cross-validation. One reports the average performance on models trained on multiple random training, validation, and test splits. This is also just what was done in \cite{varga2019no}, as the paper explains ``The different versions of our algorithm'' (different pooling strategies, different SVR kernels) ``were assessed based on KoNViD-1k by fivefold cross-validation with ten replicates in the same manner as the study by \cite{li2016spatiotemporal}.'' Checking the paper \cite{li2016spatiotemporal} confirms that  the whole dataset was split into folds, each one being used as a test set for the SVR. Therefore, 80\% of the videos contained in each such test set had already been utilized in the network fine-tuning stage. So most of the feature vectors from a test set had been learned in the feature extraction network from their corresponding video MOS values, and at the end it was the job of the SVR to predict the same MOS values from these learned features. This constitutes another clear case of data leakage resulting in `tainted' test sets, that explains why our reimplementation could not reach the performance claimed in \varga.

Based on the above analysis, we succeeded to reproduce the results published in \cite{varga2019no} with random splits into training, validation, and (tainted) test sets for the training and testing of the SVR.  On the very right of Figure~\ref{fig:results}, the average performance of 5 gaussian kernel function SVRs trained on tainted test sets is shown, with the standard deviation denoted by error bars. 


Table~\ref{tab:perf_results} provides a summary of the performance results of various VQA algorithms on KoNViD-1k. 
The middle section (rows 6 and 7) compares the original approach without fine-tuning both as reported by \varga\ and as re-computed by us, similar to the left plot of Figure~\ref{fig:results}. As described before, since no fine-tuning was performed, the data set splits have no impact, and test sets can therefore not be tainted with data items that the network had seen before. The performance numbers we obtained are very similar to those reported in \varga. 

\begin{table}
  \caption{Performance results of various VQA algorithms on KoNViD-1k. The data is taken from the references listed in the second column. The last two columns designate whether fine-tuning (column `ft') was performed correctly (green checkmark), or with data leakage (red cross), and whether the test set (column `test') was independent (green checkmark) or tainted (red cross). The approach indicated by \ssymbol{1} was published after the referenced publication and is current state-of-the-art. --.--{}-- indicates unreported values. The numbers in bold font in the last line give the true performance of the method in \varga, much below 0.85 PLCC and SROCC as claimed.}
  \label{tab:perf_results}
  \centering
  \begin{tabular}{rrccccccc}
    \toprule
        & VQA algorithm  & src & PLCC & SROCC & & &  \\
    \cmidrule{1-5}
    1 & CORNIA    & \cite{korhonen2019two} & 0.51 ($\pm$0.02) & 0.51 ($\pm$0.04)   &   &   &  \\
    2 & V-BLIINDS  & \cite{korhonen2019two} & 0.58 ($\pm$0.05) & 0.61 ($\pm$0.04)   &   &   &  \\
    3 & STFC & \cite{men2018spatiotemporal} & 0.64 ($\pm$--.--{}--) & 0.61 ($\pm$--.--{}--)   &   &   &  \\
    4 & TLVQM     & \cite{korhonen2019two}  & 0.77 ($\pm$0.02) & 0.78 ($\pm$0.02)  &   &   &  \\
    5 & MLSP-VQA-FF\ssymbol{1}    & \cite{gotz2019no}  & 0.83 ($\pm$0.02) & 0.82 ($\pm$0.02)  &   &   &  \\
    \cmidrule[1pt]{1-8}
    6 & Inception-V3 & \cite{varga2019no}  & 0.72 ($\pm$--.--{}--) & 0.68 ($\pm$--.--{}--)   & max & -  & - \\
    7 & Inception-V3   & ours & 0.73 ($\pm$0.02) & 0.70 ($\pm$0.03)    & max &  - & - \\
    \cmidrule{4-8}
    8 & Inception-V3 & \cite{varga2019no}  & 0.85 ($\pm$--.--{}--) & 0.85 ($\pm$--.--{}--)& avg &\rxmark&\rxmark \\
    9 & Inception-V3  & ours & 0.83 ($\pm$0.02) & 0.84 ($\pm$0.03)& avg &\rxmark&\rxmark \\
    10 & Inception-V3  & ours & 0.76 ($\pm$0.03) & 0.74 ($\pm$0.04)& avg &\gcmark&\rxmark \\
    11 & Inception-V3  & ours & 0.72 ($\pm$0.03) & 0.69 ($\pm$0.04)& avg &\rxmark&\gcmark \\
    12 & Inception-V3  & ours & \textbf{0.71 ($\pm$0.03)} & \textbf{0.69 ($\pm$0.04)} & avg &\gcmark&\gcmark \\
    \midrule
        &  Base architecture & src  & PLCC & SROCC & pool & ft & test \\
    \bottomrule
  \end{tabular}
\end{table}


Next, the bottom part of the table summarizes the results of the approach both as reported in \varga\ and as reimplemented by us. Here, the last two columns indicate whether fine-tuning was performed correctly (green checkmark) or with data leakage (red cross), and whether the test set was independent (green checkmark) or tainted (red cross), respectively. The reimplemented approach with incorrect fine-tuning and tainted test sets (row 9) closely matches the results reported in  \varga\ (row 8). 
The next two rows 10 and 11 show the individual impact that the two cases of data leakage have. The tainted test sets caused a larger gap in performance, which was to be expected, given that this form of data leakage is beneficial to the performance on the test set specifically. Surprisingly, the incorrect fine-tuning appears to improve results over correctly implemented fine-tuning, which deserves additional investigation.

Finally, row 12 shows the \textit{true} performance of the approach proposed in \varga. Both fine-tuning and testing were carried out correctly, with strict training, validation, and test set splitting. The average performance across five random data splits, each fine-tuned using only the training set, model selection performed using the performance on the validation set, and performance reported solely on test set items was 0.71 PLCC and 0.69 SROCC. With this result, the proposed method cannot be considered state-of-the-art, as it performs worse than TLVQM by 0.06 PLCC and 0.09 SROCC, which is a considerable performance gap. Moreover, recent advances in the field \cite{gotz2019no} have pushed performance on KoNViD-1k to 0.83 PLCC and 0.82 SROCC, shown in row 5. 

We also remark that the performance of the correctly implemented system (row 12) is worse than when not fine-tuning at all (row 7). As already described in Section~\ref{sec:finetuning}, there are concerns with the fine-tuning process that could be addressed differently, and we will discuss that in Section~\ref{sec:discussion}.

\section{Discussion}
\label{sec:discussion}
Beyond the problems with the implementation in \varga\ described above, there are some concerns with the approach in general. First, support vector machines (SVM) are not an inherently scaleable machine learning approach. Specifically, two characteristics of SVMs are problematic for scale:

\begin{itemize}
    \item The memory storage requirements for the kernel matrix of SVMs scale quadratically with the number of items and
    \item training times of traditional SVM algorithms scale superlinearly with the number of items.
\end{itemize} 

There are approaches to circumvent these problems, but for large-scale feature spaces with many data instances SVMs commonly train slower and/or perform worse than simpler approaches. The feature space of the inputs used here for VQA is close to a size that is difficult for SVMs to handle. Moreover, SVR is sensitive to model hyperparameters~\cite{tsirikoglou2017hyperparameters,ito2003optimizing}. Careful hyperparameter optimization is commonly performed to ensure robustness and reproducibility of the results.

Furthermore, it is not entirely clear, why the approach was split into two separate stages in the first place. Instead of having fine-tuned on coarse MOS classes, one could have replaced the head of the Inception-style network with a regression head, thereby eliminating the need for the SVR stage. This end-to-end training approach seems more immediate, and a comparison should have been considered, as there is potential that this streamlined approach could allow the network to leverage more information. For completeness we have evaluated this training procedure on the five random splits used throughout this article.

Following the approach of \varga, we took an Inception-V3 network, removed the layers beyond the last pooling layer, and attached three fully connected layers of sizes 1024, 512 and 32, each followed by a rectified linear unit layer that clips negative values to zero and a dropout layer with 0.25 dropout. The fully connected layers of the new head were trained at a ten times increased rate, as compared to the rest of the network. This improves the training as the weights in the layers of the head are randomly initialized, while the rest of the network is pre-trained. Lastly, we added a fully connected layer of size 1. We trained this network with stochastic gradient descent with momentum and a learning rate of $\alpha=10^{-4}$ and otherwise default training settings, except for a custom learning rate scheduler, that multiplies the learning rate by 0.75 after every epoch. The network was trained for 10 epochs total on 20\% of the frames of videos, to retain comparability to the results in Table~\ref{tab:perf_results}.

For testing, the network's prediction was computed for every frame of the test videos. A video-level score was computed as the average frame-level prediction, resulting in $0.66 (\pm 0.02)$ PLCC and $0.65 (\pm 0.03)$ SROCC. This shows, that the two-staged approach proposed in \varga\ was successful in improving video quality prediction over this na\"ive approach.

%


\section{Conclusions}
\label{sec:conclusions}
In this paper, we have tried to reproduce the performance of a machine learning approach published in \varga\ for no-reference video quality assessment. The originally reported performance numbers for the KoNViD-1k dataset were well above the state-of-the-art at the time of publication. However, our implementation of the proposed method, based mostly on the author's code, showed that the true performance is far below the claims in the paper. 


We have shown two cases of data leakage that have likely occurred in the original implementation. By introducing data leakage errors in our reimplementation, we were able to reproduce the incorrect performance values published in \varga consistently. Moreover, we brought strong arguments for the claim that the original implementation was affected by these errors, both by inspecting code published by the author and by careful examination of the description of the experimental setup.

As a complementary contribution, we evaluated an alternative direct end-to-end approach to the problem of VQA using pre-trained neural networks that should have been compared with the two-stage approach in \varga. This end-to-end approach skips the feature extraction step by immediately training a regression head.

%



%


\bibliographystyle{spmpsci}      
\bibliography{template}   

\end{document}